\documentclass[runningheads]{llncs}
\usepackage{graphicx}
\usepackage{amssymb}
\begin{document}
\title{A Game Theoretic Model for Strategic Coopetition in Business Networks}
%
%
\author{Segev Wasserkrug\inst{1} \and
Eitan Farchi\inst{1}}
\authorrunning{S. Wasserkrug et al.}
%
\institute{IBM Research, IBM Haifa Research Lab, Mount Carmel, Haifa, Israel
\email{\{segevw,farchi\}@il.ibm.com}}
\maketitle              
\begin{abstract}
Private blockchain is driving the creation of business networks, resulting in the creation of new value or new business models to the enterprises participating in the network. Such business networks form when enterprises come together to derive value through a network which is greater than the value that can be derived solely by any single company. This results in a setting that combines both competitive and cooperative behavior, and which we call strategic coopetition. Traditionally, cooperative and competitive behavior have been analyzed  separately in game theory. In this article, we provide a formal model enabling to jointly analyze these different types of behaviors and the interdependencies between them. Using this model, we formally demonstrate and analyze the incentives for both cooperative and competitive behavior.

\keywords{Game Theory  \and Business Networks \and cooperation \and competition}
\end{abstract}
\section{Introduction}
Private blockchain is driving the creation of business networks. Such networks form when enterprises come together to derive value through a network which is greater than the value that can be derived solely by any single company. In such settings, there is incentive to increase the overall value derived from the network, as this results in additional value which can be potentially divided among the network participants. However, given a fixed value derived from the network, the enterprises compete between themselves regarding the value of the network. We call settings which combine cooperative and competitive behavior in this manner \emph{strategic coopetition} settings

Traditionally, competitive and cooperative aspects have been treated separately in game theory through cooperative games and games in extensive form. Therefore, in this work, we provide a formal framework tying these two aspects together, and describe some results formalizing the connection between the two.

\subsection {\label{Example:ShippingNetwork} Example of a simple shipping network}

As an illustrating example, consider a simple shipping network composed of two shipping companies $s_1$ ans $s_2$, who can transport goods, and two \emph{cargo owners}, $c_1$ and $c_2$, whose goods are being transported. This is an example of a shipping network, such as the one supported by the Tradelens blockchain network (see https://www.tradelens.com/).

For this network, we have the following:

\begin{itemize}
    \item  $c_1$ and $c_2$ each have a single good (which we denote by $g_1$ and $g_2$ respectively), which they can sell at the respective prices of $p_{c_1}$, $p_{c_2}$.
    \item Both $c_1$ and $c_2$ need to have the item shipped by one  of the shippers $s_1$ or $s_2$.
    \item Both $s_1$, $s_2$ have a cost $p_{s_1}$, $p_{s_2}$ respectively for shipping any single good.
\end{itemize} 

An example possible outcome is that $c_1$ ships the good with $s_1$ and $c_2$ ships the good with $s_2$. In order to have the good shipped, each carrier $c_i$ for  $i \in \{1,2\}$ would have to pay a shipping cost $v_{(i,i)}$ such that $p_{s_i}<v_{(i,i)}<p_{c_i}$ for $i \in \{1,2\}$. Clearly, without cooperation between the shippers and the cargo owners, the value to all participants is zero.

This simple setting demonstrates some aspects of such strategic coopetition settings:
\begin{itemize}
    \item The value is derived from a flow of \emph{goods} between the parties.
    \item There are costs and benefits external to the parties in the network which, when flowing between the network participants, provide value to the participants.
\end{itemize}

\section{Formal Model}

To formally capture such strategic coopetition settings in business networks we define the following three components:
\begin{enumerate}
    \item The players are a set of $n$ \emph{enterprises} or
    \emph{companies} participating in the network. We denote this set by $C=\{1,\ldots,n\}$.
    \item The flow of \emph{goods} between these enterprises.
    \item The flow of \emph{value} or utility between these enterprises, that is provided in exchange for the flow of goods.  As the flow of goods between one enterprise to another corresponds to the flow of utility between these enterprises and is associated with some underlying agreements between them we assume the possibility of side payments.   
\end{enumerate}

Below, we formally define the flow of goods and values.

\subsection{Flow of goods}

The flow of goods for each company $i \in C$ is formalized as follows:

\begin{itemize}
    \item  $C$ has a  set of $j_i$ good types $G_i=\{G_i^1,\ldots,G_i^{j_i}\}$ which it can transfer to other enterprises. We will denote all of the good types of all companies by $G=\cup_i G_i$.
    \item  The goods received by company $i$ from other companies are transformed into a different set of goods that $i$ can  transfer. In order to model such transformations, the company $i$ has a  transformation function $t_i: N^{G} \rightarrow N^{G}$ where $t_i(g)$ $k$'th element is the number of goods of type $k$ in $t_i(g)$, $k \in G$.  This function  indicates for each vector of input goods $g \in N^{|G|}$ the vector of output goods produced, $t_{i}(g)$, that $i$ can then transfer to other companies.  Naturally, $t_i(g)_j > 0$ iff $j \in G_i$ as we assume company $i$ can only transform goods of type in $G_i$ to other companies. 
\end{itemize}

\subsection{Flow of value}

The flow of value is used to define the utility of each player in the game. It is strongly tied to the flow of goods. As the source of value for companies in the network results from the transfer of goods into and out of the network,  for each enterprise, $i$, we define two functions that help formalize the flow of value through the network as follows.   

\begin{enumerate}
    \item Let $P_i = \{g = (g_i)_{i \in G} \in N^G | g_i > 0$ only if $ i \in G_i \}$ be the set of goods a company may produce.  Define an \emph{external benefit} function $eb_i: P_i \rightarrow \mathbb{R}_{\geq 0}$.  $eb_i(g)$ defines the external benefit company $i$ obtains from transferring $g$ of goods outside the system. 
    \item Let $GT_i = \{ (g, g') \in N^G \times N^G | g'=t_i(g) \}$.  $GT_i$ is a set of pairs of input goods to $i$, $g \in N^G$, and the goods produced by these input goods, $g' = t_i(g)$.  Define an external cost function, $ect_i : GT_i \rightarrow \mathbb{R}_{\geq 0}$.   $ect_i(g, g')$ represents the cost of transforming the input goods $g$ into the output goods $g'$, or from obtaining such goods from a source external to the network.
\end{enumerate}

\subsection{Business network games}
\label{graph}

Based on the above definitions, and to enable analyzing such settings, we define a class of games which we call \emph{business network games}. A business network game is any game whose outcomes can be defined by a graph as follows.

\begin{itemize}
    \item The nodes in the graph $V$ are $\{C\} \cup \{s\} \cup \{t\}$ where $s$ is a special source node and $t$ is a special sink node.
    \item There is directed edge (i,j) between two nodes $i,j \in C$ such that $i$ and $j$ have agreed on a transfer of $\bar{g}_{i,j}$ from $i$ to $j$ for a value of $v_{i,j}$. In more details, $i$ has offered to transfer the set of goods $\bar{g}_{i, j}$ for the value $v_{i,j}$ and $j$ has offered to purchase the goods $\bar{g}_{i, j}$ for the value of $v_{i,j}$. On edge $(i,j)$ there will be a flow of goods $\bar{g}_{i, j}$ from $i$ to $j$ and a flow of the value $v_{i,j}$ from $j$ to $i$.  We will refer to $v_{i, j}$ as an internal benefit for $i$ and an internal cost for $j$. 
    \item For every outside transfer of goods $\bar{g}_i \in P_i$ from a node $i \in C$ there will be an edge $(i,t)$. On this edge, there will be a flow of $\bar{g}_i$ goods from $i$ to $t$ and a flow of $eb_i(\bar{g_i})$ value from $t$ to $i$ (we will also denote this value by $v_{i,t}$).
    \item Each node $i \in C$ must maintain \emph{goods flow conservation}. I.e, let $\bar{g}_{in}$ be the vector of the sum of all goods flowing into $i$ and $\bar{g}_{out}$ the vector of the sum of all goods on the edges flowing out of $i$. Then it must hold that $t_i(\bar{g}_{in})\leq \bar{g}_{out}$. 
    \item For every $i \in C$ there is an edge $(s,i)$ and the value flow on the edge is $ec_i(\bar{g}_{in},t_i(\bar{g}_{in}))$. We will also denote this value by $v_{s,i}$.  In addition, we assume that there may be some flow of goods $i$ is getting "for free".  For example, $i$ may be mining some natural resource in its position.  For free means here that the resource is obtained from outside the network and is not paid by $i$.  We will model this as a vector of goods $\bar{g}_{s, i} \in N^G$ associated with $(s,i)$. 
\end{itemize}

For any company $i \in C$, the utility of the game for player $i$ is defined as $\Pi_i=\Sigma_{(i,j), j \in C}v_{i,j}-\Sigma_{(j',i), j' \in C}v_{j',i}$

We also define the concept of \emph{Total Network Value}, or $TNV$, to be value of the network. The $TNV$ is defined as $TNV=\Sigma_{(i,t), i \in C}v_{i,t}-\Sigma_{(s,i), i \in C}v_{s,i}$.

In the context of the simple example from section \ref{Example:ShippingNetwork} with the agreements described between $c_1$ and $s_1$ and between $c_2$, and $s_2$, $\Pi_{s_i}=v_{i-i}$, $\Pi_{c_i}=p_{c_i}-v_{(i,i)}$, and $TNV=p_{c_1}+p_{c_2}-(p_{s_1}+p{s_2})$

\section{Relation between network value and individual value}

An important property of such games is that in any such outcome as described by a graph defined in section \ref{graph}, it holds that $\Sigma \Pi_i=TNV$. This can be proven by looking at the edges appearing in  $\Sigma \Pi_i$, as:
\begin{itemize}
    \item Every edge $(i,j)$ such that $i,j \in C$ appears twice: once with a positive sign in $\Pi_i$ and once with a negative sign in $\Pi_j$. Therefore, the sum of all such edges is zero.
    \item Every edge $(i,t)$ for $i \in C$ appears once with a positive sign in $\Pi_i$
    \item Every edge $(s,i)$ for $i \in C$ appears once with a negative sign in $\Pi_i$
\end{itemize}

This result highlights an interesting relations between cooperation and competition as increasing the $TNV$ (the right hand side of the above equation) automatically increases the overall sum of values for all of the participants in the network. However, it could also be the case that such an increase is a result of an increase of the values of some participants while lowering, not by the same amount, the values of the others. However, reducing the value of any single participant in the network may cause that participant to leave the network. Therefore, it is desirable to show an additional property, namely, that for any increase in $TNV$, there exists at least one flow for which all parties enjoy increased value. This property is stated formally in the following lemma.

\begin{lemma}
Let there be a setting in which there is a valid flow of goods and value, and let $TNV_1$ be its corresponding $TNV$. If there exists a valid flow of goods on the network $TNV_2=\Sigma_{(i,t), i \in C}v_{i,t}-\Sigma_{(s,i), i \in C}v_{s,i}$ is such that $TNV_2>TNV_1$, then there exists a valid flow of value in the network so that $\forall i \in C, \Pi_i^2>\Pi_i^1$. (Where $\Pi_i^j$ for $j \in \{1,2\}$ is respectively the value of $i$ for the respective goods flow.)
\end{lemma}

Below is the proof sketch, which is a proof by induction on the number of participants in the network:
\begin{itemize}
    \item For the base of the induction, the single node, this property holds by definition.
    \item for any network with $n+1$ nodes, we select arbitrarily two nodes and collapse them into a single node, altering their $t$, $eb$, and $ec$ functions to maintain the flow properties for both goods and values.
    \item On the network with $n$ nodes, we have increase value flow for all nodes, which for the combined node is greater than the sum of values of the individual nodes in the original network
    \item We split back the nodes, maintaining the goods flow and splitting the value flow.
\end{itemize}  

Note, however, that finding such a flow may decouple the goods flow from the value flow. For example, assume that in the simple network example, $p_{s_2}>p_{s_1}$ the $TNV$ can be increased by $c_2$ shifting to also shipping the good through $s_1$. However, this means that, in order to ensure that the $\Pi_{s_2}$ in the new flow is increased, $s_2$ is getting paid even without shipping any good. Note that while paying $s_2$ for not shipping any good is counterintuitive, there are justifications for doing this. With zero value, $s_2$ may decide to leave the network. If this occurs, $s_1$ will have a monopoly on shipping, enabling $s_1$ to significantly increase its shipping price.

\section{Related work}
There are quite a few examples of business networks that have been analyzed through game theory. One example is a simple supplier and retailer networks such as the one appearing in \cite{ErtogralBargaining2001}. There are also some works that explicitly address coopetitive games (see, for example, \cite{Carfi2015}). However, to the best of our knowledge, no previous works have explicitly characterized the relation between the cooperative and competitive aspects appearing in such networks.

Another related question are what are the specific games that can lead to the creation of such networks. In \cite{ErtogralBargaining2001}, a bargaining game is defined over a simple network composed of a single buyer and multiple suppliers. Defining such games is one of our current work focus, and is closely related to the topic of stability in trade networks (see \cite{ostrovsky2008stability} and \cite{hatfield2015chain})

\section{Summary and future work}
In this work, we have provided a formal definition of a setting we have termed strategic coopetition which appears in business networks, including ones based on private blockchain. We have shown the connection between the cooperative and competitive incentives of such networks through the notion of a total network value, or $TNV$, and its strong connection to the values of the individual participants. We have also shown that an increase in the $TNV$ can result in increased value to all participants.

Future directions we are now focusing on are the creation of distributed algorithms and explicit games for establishing such networks, and increasing both network and individual value. We believe that such algorithms could significantly accelerate the value provided by private blockchain networks.

%
%
%
\bibliographystyle{splncs04}
\bibliography{blockchainws.bib}

\begin{thebibliography}{1}
\providecommand{\url}[1]{\texttt{#1}}
\providecommand{\urlprefix}{URL }
\providecommand{\doi}[1]{https://doi.org/#1}

\bibitem{Carfi2015}
Carfì, D.: A model for coopetitive games (06 2015).
  \doi{10.13140/RG.2.1.2929.0960}

\bibitem{ErtogralBargaining2001}
Ertogral, K., Wu, S.: A bargaining game for supply chain contracting  (01 2001)

\bibitem{hatfield2015chain}
Hatfield, J.W., Kominers, S.D., Nichifor, A., Ostrovsky, M., Westkamp, A.:
  Chain stability in trading networks. Available at SSRN 3180740  (2015)

\bibitem{ostrovsky2008stability}
Ostrovsky, M.: Stability in supply chain networks. American Economic Review
  \textbf{98}(3),  897--923 (2008)

\end{thebibliography}
%




\end{document}